\documentclass[journal=jacsat,manuscript=article,layout=twocolumn]{achemso}
\usepackage{graphicx}
\usepackage{amssymb,amsmath}

\author{Mie Andersen}
\email{mie.andersen@ch.tum.de}
\affiliation{Chair for Theoretical Chemistry and Catalysis Research Center, Technische Universit{\"a}t M{\"u}nchen, Lichtenbergstr. 4, 85747 Garching, Germany}
\author{Sergey V. Levchenko}
\affiliation{Fritz-Haber-Institut der Max-Planck-Gesellschaft, Faradayweg 4-6, 14195 Berlin, Germany}
\author{Matthias Scheffler}
\affiliation{Fritz-Haber-Institut der Max-Planck-Gesellschaft, Faradayweg 4-6, 14195 Berlin, Germany}
\author{Karsten Reuter}
\affiliation{Chair for Theoretical Chemistry and Catalysis Research Center, Technische Universit{\"a}t M{\"u}nchen, Lichtenbergstr. 4, 85747 Garching, Germany}

\title{Beyond scaling relations for the description of catalytic materials}

\begin{document}

\begin{tocentry}
\includegraphics[width=0.75\textwidth]{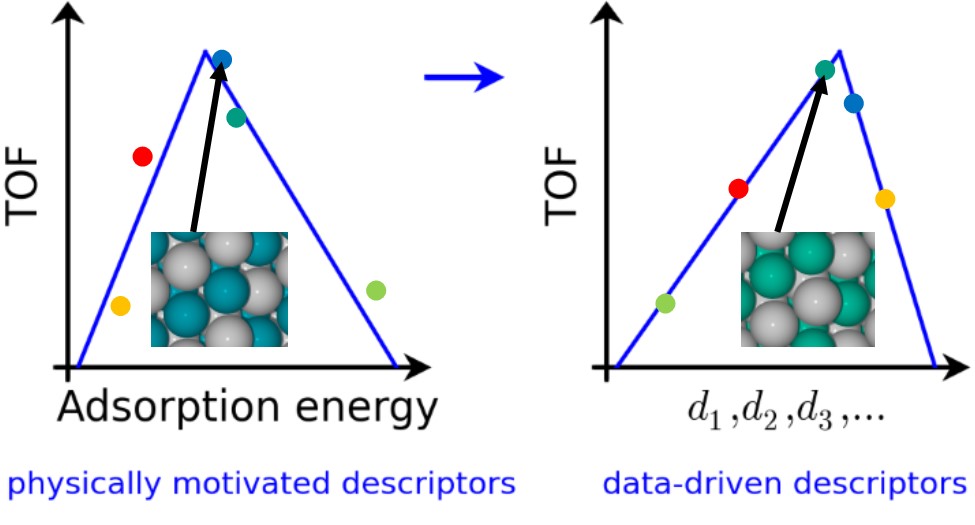}
\end{tocentry}

\begin{abstract}
Computational screening for new and improved catalyst materials relies on accurate and low-cost predictions of key parameters such as adsorption energies. Here, we use recently developed compressed sensing methods to identify descriptors whose predictive power extends over a wide range of adsorbates, multi-metallic transition metal surfaces and facets. The descriptors are expressed as non-linear functions of intrinsic properties of the clean catalyst surface, e.g.\ coordination numbers, $d$-band moments and density of states at the Fermi level. From a single density-functional theory calculation of these properties, we predict adsorption energies at all potential surface sites, and thereby also the most stable geometry. Compared to previous approaches such as scaling relations, we find our approach to be both more general and more accurate for the prediction of adsorption energies on alloys with mixed-metal surfaces, already when based on training data including only pure metals. This accuracy can be systematically improved by adding also alloy adsorption energies to the training data.
\end{abstract}

\section{Introduction}
Surface catalysis has an enormous impact on the environment and our society's health and prosperity \cite{Ertl2008}. However, the reliable description of catalytic properties and the prediction of what materials may be even better catalysts than what we know today is still weak. 
This is due to the highly nonlinear and intricate relationship between the "catalyst material" (the static material that is introduced before the catalytic process is running) and the strongly kinetically controlled surface reactions at realistic conditions \cite{Reuter2004,Reuter2016}. Simply speaking, the basic understanding of heterogeneous catalysis is given by the Sabatier principle and the Br{\o}nsted-Evans-Polanyi (BEP) relation \cite{Norskov2009}. The first states that there is an optimum adsorption strength for which the reactants bind strong enough to allow for adsorption and dissociation into reaction intermediates, but weak enough to allow for consecutive desorption of products. In turn, the BEP concept tells that the energy barriers of the chemical reactions scale approximately linearly with the adsorption energies of the molecules. 

In consequence, the reliable prediction of adsorption energies is a key element of any theoretical description and search for new catalyst materials. For this, we here present a data-driven approach that does not start from a specific physical model, e.g. the tight-binding description of chemical bonding, but accepts that the intricacy of processes that cooperate or compete in materials properties may not necessarily be describable by a closed physical equation. This has been described as the fourth paradigm of materials science \cite{Draxl2018}. Previous data-driven approaches to the prediction of adsorption energies \cite{Ma2015,Li2017,Ulissi2017,Gasper2017,Jinnouchi2017} have exploited an approximately linear correlation between the adsorption energies of certain adsorbates on pure transition metal (TM) surfaces (scaling relations \cite{Abild2007}) to extend the data-driven predictions of one or two species to other adsorbates involved in the reaction. We instead directly learn the adsorption energies of a whole range of atoms and molecules at all potential adsorption sites (thereby also the most stable site) only from properties of the clean surface. This allows us to go beyond scaling relations and the often unfulfilled assumptions tied to this particular physical model. It furthermore opens the perspective of directly searching for outliers to scaling relations, which can be highly interesting catalyst materials missed by the standard approach \cite{Grabow2012,Kyriakou2012,Darby2018}.

The method for identifying the key descriptive parameters is the recently developed compressed sensing method SISSO \cite{Ouyang2017} (sure independence screening and sparsifying operator), which enables us to identify the best multidimensional descriptor out of an immensity of candidates (billions). Our descriptors are more general and less costly to use than previous approaches and allow for making predictions for a huge number of surfaces including both multi-metallics and various facets. Through BEP relations our approach can also describe chemical reactions, as well as the diffusion of atoms and molecules at the surface visiting metastable adsorption sites.

\section{Computational details}
\subsection{Density Functional Theory}
\begin{figure}
\centering
\includegraphics[width=\columnwidth]{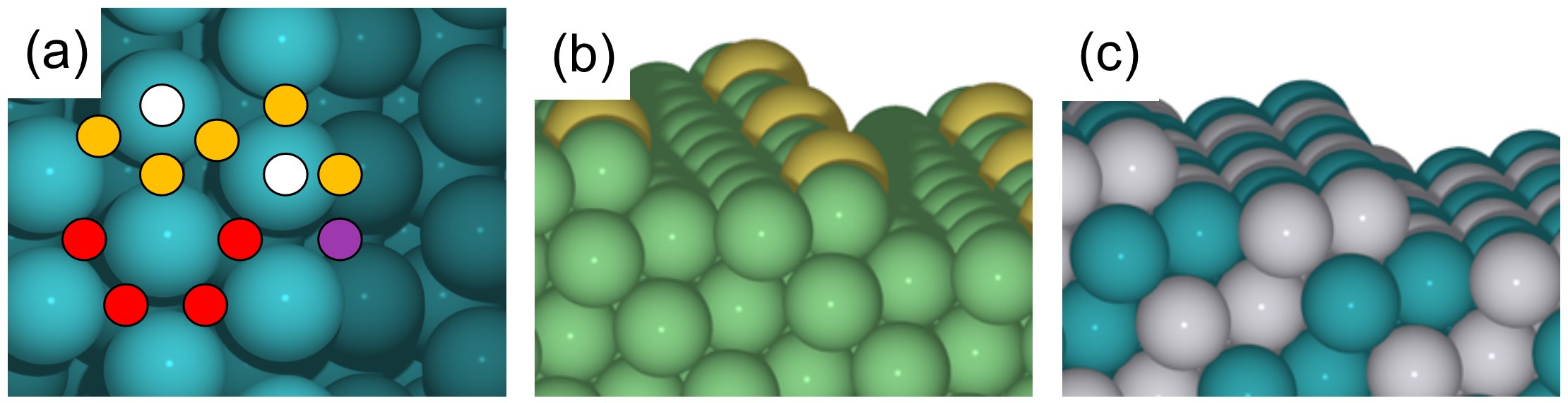}
\caption{Top view of the structure of (a) the fcc(211) facet along with the considered terrace and step adsorption sites, which cover one four-fold-coordinated site (purple dot), four three-fold-coordinated (fcc and hcp) sites (red dots), five bridge sites (yellow dots) and two top sites (white dots). Perspective views of the structures of (b) a single-atom (SA) alloy and (c) an AB alloy.}
\label{fig:structures}
\end{figure}

The data sets employed in the present work were obtained from plane-wave density-functional theory (DFT) calculations (Quantum ESPRESSO code \cite{Giannozzi2009}) using the van der Waals-corrected BEEF-vdW exchange-correlation functional \cite{Wellendorff2012}. The larger data set consists of adsorption energies of atomic and molecular adsorbates (C, CH, CO, H, O, OH) on the stepped fcc(211) facets of nine TMs (Ni, Cu, Ru, Rh, Pd, Ag, Ir, Pt and Au) and selected single-atom (SA) and AB bimetallic alloys. The metals were modeled in fcc stacking using a $(1\times3)$ ($(1\times2)$ for AB alloys) supercell with fifteen metal layers. The considered adsorption sites are illustrated in Fig.\ \ref{fig:structures}(a) and cover both high-symmetry terrace and step sites. For each adsorbate, all adsorption sites that correspond to local minima on the potential energy surface were included. The SA alloys \cite{Han2009,Kyriakou2012,Darby2018} were constructed by replacing one metal atom at the step with a different metal (see Fig.\ \ref{fig:structures}(b)). Specifically, we considered Ag@Cu (Ag atom in Cu surface), Pt@Rh, Pd@Ir and Au@Ni. For the AB L1$_0$ alloys \cite{Curtarolo2012} (AgPd, IrRu, PtRh and AgAu) the considered surface termination is depicted in Fig.\ \ref{fig:structures}(c). The total numbers of adsorption energies in the data set are 344 (metals), 281 (SA alloys), and 259 (AB alloys). The smaller facets data set consists of one adsorption site for each adsorbate on the fcc(111), (110) and (100) facets of the nine TMs, leading to a total of 54 adsorption energies on each facet. All data sets are compiled in Supplementary Sec.\ S1 together with further computational details.

\subsection{Compressed sensing}
The SISSO method \cite{Ouyang2017} employed for descriptor identification makes the ansatz that the properties of interest $P^j_1,...,P^j_N \in \mathbb{R}$, (in this case a vector of $N$ adsorption energies of adsorbate $j$) can be expressed as linear functions of candidate features ${\mathbf d}_1,...,{\mathbf d}_M \in \mathbb{R}^N$, where the features are constructed as non-linear functions of user-defined primary features (see below). SISSO identifies the few best features (the number of which corresponds to the dimensionality of the descriptor) out of immense feature spaces by use of the sparsifying $\ell_0$ constraint. This is carried out in a smaller feature subspace selected by a screening procedure (sure independence screening (SIS)). The size of the subspace is equal to a user-defined SIS value times the dimension of the descriptor. 

In this work we make use of multi-task learning \cite{Ouyang2018} to identify common descriptors for the adsorption energies of several different adsorbates simultaneously. That is, the identified features are constrained to be identical for every adsorbate, while the fitting coefficients are allowed to vary between the adsorbates. We find that multi-task learning gives a better predictive performance compared to the identification of separate descriptors for each adsorbate (see Supplementary Fig.\ S2). We consider two hyperparameters in the SISSO method: the dimension of the descriptor and the feature space rung (see below) as well as the SIS value (see Supplementary Sec.\ S2), that we fix for the current application through a validation data set (see below).

\subsection{Primary features}
The decision which primary features to use as input for the feature construction is crucial for the predictive performance of the resulting descriptors. Inspired by previous studies \cite{Fukui1982,Yang1985,Hammer1995,Ruban1997,Xin2012,Xin2014,Calle-Vallejo2015,Ma2015,Li2017} we consider four classes of primary features (see Table \ref{Tab:PFs}) related to the metal atom, the metal bulk, the metal surface and the metal adsorption site. For pure metals the primary features of the site class were calculated as averages over the metal atoms making up the site ensemble, while for alloys this was the case for the primary features of all classes. We note that the consideration of fixed adsorption sites as well as the averaging over the site ensemble is an approximation. It may break down in case of surface reconstruction or any other appearance of new adsorption motifs that were not accounted for in the calculation of the primary features. Further details regarding the primary features and all data are given in Supplementary Sec.\ S1.

\begin{table}
\resizebox{\columnwidth}{!}{%
\caption{\label{Tab:PFs}Primary features used for the feature construction.}
    \begin{tabular}{ @{\extracolsep{\fill}}l  l  c}
		  \hline\hline
      Class & Name & Abbreviation \\
			\hline
			Atomic & Pauling electronegativity & PE \\
			       & Ionization potential & IP \\
			       & Electron affinity & EA \\
			\hline
		  Bulk & fcc nearest neighbor distance & bulk$_{\rm nnd}$ \\
			     & Radius of $d$-orbitals & $r_d$ \\
					 & Coupling matrix element squared & $V_{\rm ad}^2$ \\
			\hline
			Surface & Work function & $W$ \\
			\hline
			Site          & Number of atoms in ensemble & site$_{\rm no}$ \\
										& Coordination number & CN \\
										& Nearest neighbor distance & site$_{\rm nnd}$ \\
			              & $d$-band center & $\varepsilon_d$ \\
			              & $d$-band width & $W_d$ \\
										& $d$-band skewness & $S_d$ \\
										& $d$-band kurtosis & $K_d$ \\
										& $d$-band filling & $f_d$ \\
										& $sp$-band filling & $f_{sp}$ \\
										& Density of $d$-states at Fermi level & DOS$_d$ \\
										& Density of $sp$-states at Fermi level & DOS$_{sp}$ \\
			\hline\hline
    \end{tabular}}
\end{table}

\subsection{Feature construction}
As discussed above, candidate features are constructed as non-linear functions of the primary features. In the SISSO method this is achieved in practice by applying algebraic/functional operators such as addition, multiplication, exponentials, powers, roots etc.\ to the features \cite{Ouyang2017}. A full list of the used operators can be found in Supplementary Sec.\ S2. Arbitrarily large feature spaces can be constructed by iteratively applying these operators to the already generated features. The starting point $\mathbf{\Phi}_0$ corresponds to the 18 primary features listed in Table \ref{Tab:PFs}. We consider up to three iterations, generating thereby the feature spaces $\mathbf{\Phi}_1$, $\mathbf{\Phi}_2$ and $\mathbf{\Phi}_3$. Note that a given feature space $\mathbf{\Phi}_n$ contains also all of the lower rung feature spaces. The $\mathbf{\Phi}_1$ and $\mathbf{\Phi}_2$ feature spaces are still comparatively small. They consist of 783 and about 10$^6$ features, respectively. For the third iteration generating $\mathbf{\Phi}_3$ the approach we chose consisted in carrying out two rounds of feature construction and descriptor identification, each for a subset of only 16 out of the 18 primary features, in order to limit the $\mathbf{\Phi}_3$ feature space of each round to a tractable value of about $10^{11}$. In the first round the skewness and kurtosis of the $d$-band were excluded, since the higher order $d$-band moments are expected to be less important than the lower order moments. Among the identified best descriptors (i.e.\ with lowest validation errors, see below) of the first round, two primary features of the site class never appeared, namely the nearest neighbor distance and the density of $d$-states at the Fermi level. In the second round these two primary features were then excluded, while the skewness and kurtosis of the $d$-band were re-included. At every dimension the best performing $\mathbf{\Phi}_3$ descriptor originated from the first round and therefore only the results of the first round are presented below.

\section{Results and discussion}
\subsection{Scaling relations}
\begin{figure*}
\centering
\includegraphics[width=\textwidth]{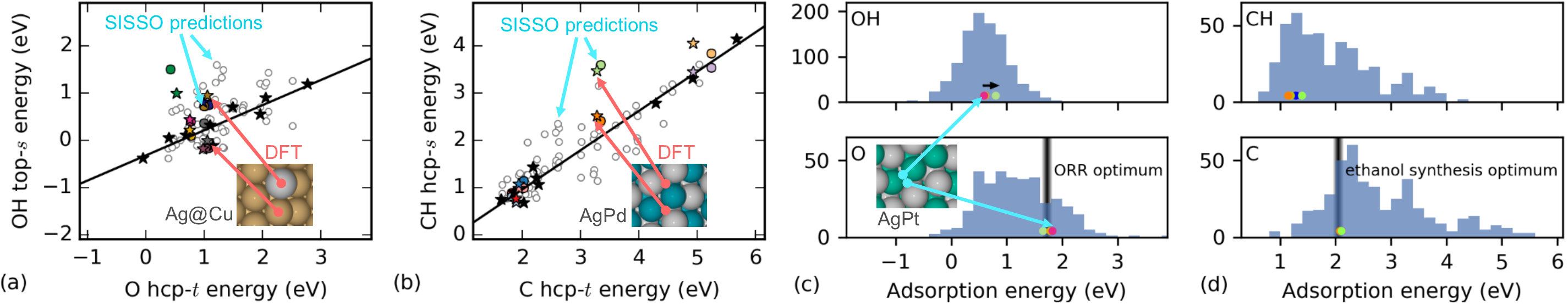}
\caption{Scaling relations for adsorption energies of (a) OH at the top-$s$ site and (b) CH at the hcp-$s$ site. The black lines are fits to the DFT adsorption energies on the pure metals (black stars). Explicitly calculated DFT adsorption energies for (a) SA alloys and (b) AB alloys are shown with colored stars. Some particularly large deviations between predictions from scaling relations and actual adsorption energies for the alloys are highlighted. SISSO predictions (8D, $\mathbf{\Phi}_3$ descriptor trained on the pooled metals and alloys data set, see text) for the calculated alloys (32 additional AB alloys) are shown with colored (gray) circles. For the prediction of the DFT-calculated alloys, these predicted data points were excluded from the training set. All scaling relations can be found in Supplementary Fig.\ S3. Histograms of SISSO-predicted adsorption energies on all potential adsorption sites of all 36 AB alloys for (c) O and OH and (d) C and CH. The black shaded regions highlight literature “volcano optimal” adsorption energies for the oxygen reduction reaction (ORR) on (111) facets \cite{Norskov2004} and for selective ethanol synthesis on (211) facets \cite{Medford2014b}. The colored circles mark those materials for which the predicted most stable (c) O adsorption energy among the (111)-like (terrace) sites and (d) C adsorption energy among all (211) sites falls within the desired range. The corresponding most stable OH and CH adsorption energies for these materials are also marked in the histograms above. A predicted near-optimal ORR material (AgPt) that breaks the O-OH scaling relation due to different metal compositions of the preferred O and OH adsorption sites is highlighted. The black arrow points from the SISSO-predicted to the scaling-relation-predicted OH adsorption energy.
}
\label{fig:scaling}
\end{figure*}

We begin by evaluating the performance of prevalent scaling relations for predicting adsorption energies on SA and AB alloys. In Fig.\ \ref{fig:scaling}(a) and (b) we show two examples of scaling relations constructed by linear fits to the DFT-calculated adsorption energies on the pure TMs (black stars). Corresponding explicitly calculated adsorption energies on SA and AB alloys are also indicated by colored stars. While many bimetallics are well described by the linear scaling relations, there are also a number of serious outliers. Some systems with particularly large prediction errors of the order of 1\,eV are highlighted. They typically contain mixed-metal sites made up of metals with very different reactivity towards O (e.g.\ Cu and Ag) or C (e.g.\ Ag and Pd). This poor performance of scaling relations derives from their calculation of the descriptors at one specific site, which fails to account for the variation in metal composition of the many other sites on the alloy surface. This issue likely occurs most severely for the considered thermochemical scaling relations. BEP relations for activation energies, in contrast, are more local in the sense that often both the transition state and the initial and final reaction intermediates coordinate (or can be chosen to coordinate) to the same metal atoms at the considered site ensemble. Correspondingly, BEP relations are typically found to exhibit significantly lower errors than thermochemical scaling relations even for the pure metals \cite{Andersen2017}.

We note that an alternative scaling-relation-based approach is to calculate all potential adsorption sites for the descriptors on a mixed-metal surface and then to consider only the most stable adsorption sites \cite{Studt2012}. However, at concomitantly increased screening costs this still does not alleviate the problem since different adsorbates (e.g.\ O and OH) generally adsorb to different site types (e.g.\ O typically prefers higher coordinated sites than OH). As a consequence, the metal composition of the preferred sites could be different.
We will come back to this point in the discussion of the compressed sensing results. In addition, not only the most stable sites, but also metastable sites missed by this approach, can get populated at higher coverages and then play an important role in the catalytic pathway \cite{Andersen2017}.

\subsection{Descriptor identification}
\begin{figure}
\centering
\includegraphics[width=\columnwidth]{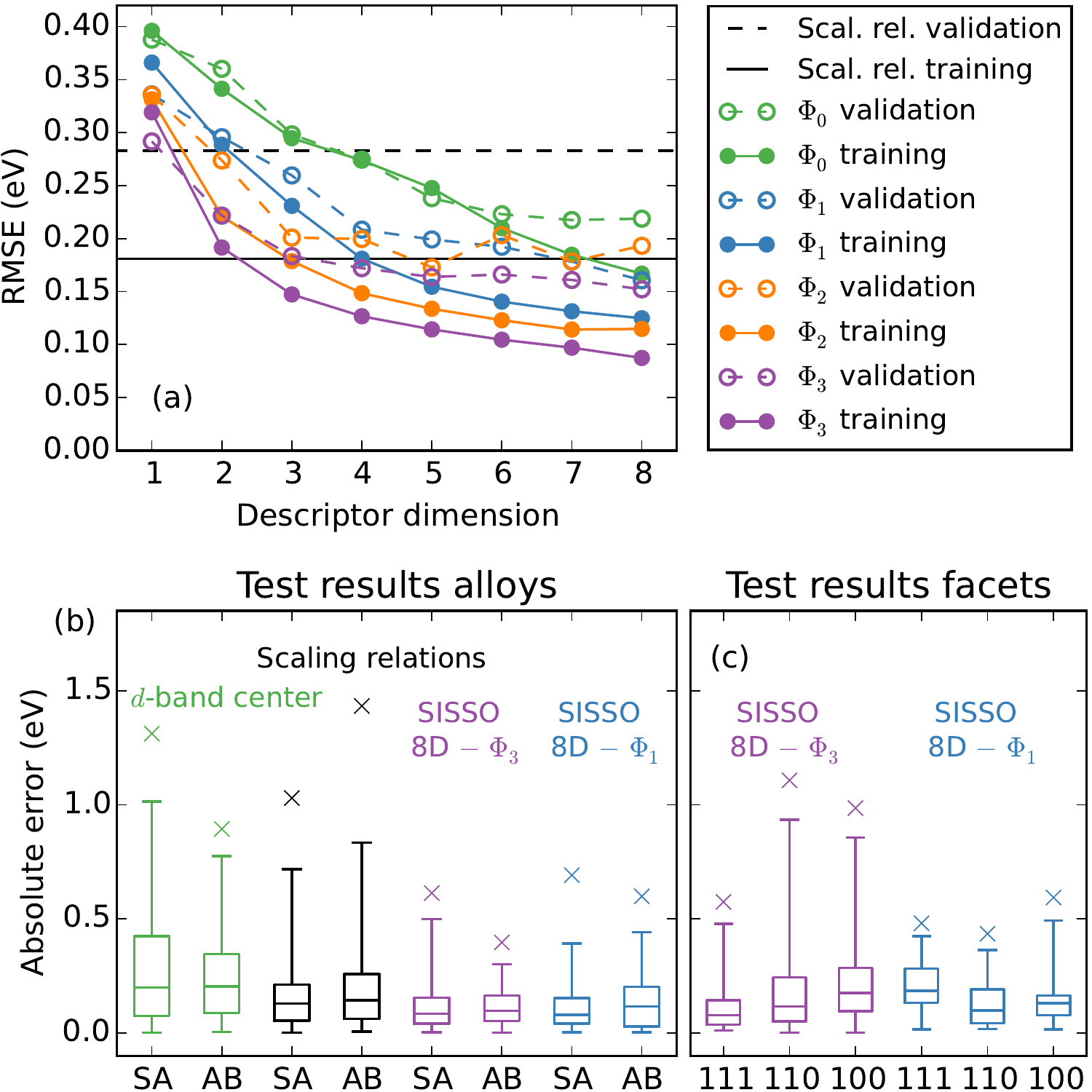}
\caption{(a) RMSE for the descriptors identified using exclusively the pure metals data set for training and 50\% of the alloys data for validation as well as corresponding results for scaling relations. (b) Box plots of the absolute errors on the test set consisting of the remaining 50\% of the single-atom (SA) and AB alloys data for the $d$-band center, for scaling relations, for the best SISSO descriptor for the validation data set (8D, $\mathbf{\Phi}_3$) and for the best SISSO descriptor when including 50\% of the (111), (110) and (100) facets data set in the validation data (8D, $\mathbf{\Phi}_1$, see Supplementary Fig. S5). The upper and lower limits of the rectangles mark the 75\% and 25\% percentiles, the internal horizontal line marks the median, and the "error bars" mark the 99\% and 1\% percentiles. The crosses mark the maximum absolute errors. (c) Corresponding box plots for the two SISSO descriptors on the facets data sets, where for the (8D, $\mathbf{\Phi}_1$) descriptor only the remaining 50\% of the facets data set not used for validation are included.}
\label{fig:metals_RMSE}
\end{figure}

The demonstrated failure of scaling relations to predict accurate adsorption energies on alloys with mixed-metal surfaces, as well as the high cost associated with the calculation of two or more adsorption energies on each alloy to be screened, emphasizes the need for new, accurate and low-cost descriptors for computational screening. In Fig.\ \ref{fig:metals_RMSE}(a) we compare the performance of scaling relations to new descriptors identified by SISSO in terms of the root-mean-square error (RMSE) on training and validation data sets. We define the best descriptor as the descriptor that achieves the lowest RMSE on the validation data set. In the calculation of the RMSE the same weight is given to every adsorbate considered in the multi-task learning irrespective of how many data points exist for the adsorbate (see Supplementary Table S1). The training data consists exclusively of the adsorption energies on the {\em pure} metals and the validation data consists of 50\% of each of the SA and AB alloys data. SISSO data are shown for 1D-8D descriptors identified from each considered feature space $\mathbf{\Phi}_n$. For each case a number of SIS values have also been tested (see Supplementary Fig.\ S4) and the best descriptor (i.e. the descriptor with the lowest RMSE on the validation data set) is shown. As expected, the SISSO training errors systematically decrease when increasing either the complexity and size of the feature space (larger $n$) or the dimensionality of the descriptor. The validation errors show the same trend, but the errors level out around the 5D to 8D descriptor depending on the rung of the used feature space. We would expect the validation errors to increase again at even higher dimensions. However, such higher dimensions are outside of the scope of the present study since the leveling out of the validation errors suggests that going beyond 8D is unlikely to result in descriptors with lower validation errors.

The 5D to 8D descriptors of $\mathbf{\Phi}_3$ all have very low validation errors, differing from each other by only about ten meV. Likely, there is no statistical significant difference in their performance. A detailed statistical analysis to derive error bars is outside of the scope of this work, which aims at a simple comparison to scaling relations. The latter are usually derived based on only one fixed training data set considering only the pure metals and we therefore follow the same approach here. In the absence of error bars we choose the descriptor with the lowest observed validation RMSE (of 0.15\,eV) as our best descriptor, i.e.\ the 8D, $\mathbf{\Phi}_3$ descriptor. This descriptor is significantly better than scaling relations, for which the validation RMSE is 0.28\,eV (horizontal dashed line). In fact, already the 2D descriptor of $\mathbf{\Phi}_3$ (with a validation RMSE of 0.22\,eV) performs better than scaling relations. The best descriptor among the primary features (the SISSO 1D, $\mathbf{\Phi}_0$ descriptor) is found to be the $d$-band center, i.e. SISSO identifies the physics that has already been discovered in form of the $d$-band model more than twenty years ago \cite{Hammer1995}.

A comparison of the performance of the $d$-band center, scaling relations and the best (8D, $\mathbf{\Phi}_3$) SISSO descriptor on the test data set (the remaining 50\% of the SA and AB alloys data) is shown in Fig.\ \ref{fig:metals_RMSE}(b). The RMSEs are: $d$-band center: 0.37\,eV, scaling relations: 0.28\,eV, SISSO: 0.15\,eV, and the maximum absolute errors (maxAEs) are: $d$-band center: 1.31\,eV, scaling relations: 1.43\,eV, SISSO: 0.61\,eV. Here, the maxAE is the maximal error observed for any of the adsorbates considered in the multi-task learning. As already observed for the validation data, the test results thus demonstrate the great improvement of the new SISSO descriptor compared to previous approaches. In Supplementary Table S7 we provide additional information about the largest deviations between calculated and SISSO-predicted adsorption energies for the alloys. In general, the C adsorption energies are the most difficult to predict since they vary much more over the TM series than e.g.\ the H adsorption energies. In addition, the most difficult alloys to predict adsorption energies for are those that combine a more noble and a more reactive metal such as Au@Ni or Ag@Cu. The maximum absolute error of 0.61\,eV is found for the OH adsorption energy on top of the Au atom in the Au@Ni alloy. The embedding of the larger and more noble Au atom in the smaller lattice constant and more reactive Ni surface (see Fig.\ \ref{fig:structures}(b)) probably provides an adsorption site that is both geometrically and electronically very different from everything else in the data set and thus harder to predict.

\subsection{Transferability of descriptors}
To further test the predictive performance of the best (8D, $\mathbf{\Phi}_3$) SISSO descriptor we show in Fig.\ \ref{fig:metals_RMSE}(c) the error distribution for the prediction of adsorption energies on three new facets, the (111), (110), and (100) facets. Some sites, in particular those found on the (111) facet, are very similar to sites found on the (211) facet in the training data. However, the (110) and (100) facets contain sites that are very different, albeit of similar coordination numbers (7 -- 9.5), compared to the (211) facet sites. It is seen that the descriptor with the best performance for (211) facets performs very well for the (111) facet (RMSE of 0.17\,eV), but significantly worse for the (110) and (100) facets (RMSEs of 0.29\,eV and 0.30\,eV, respectively). This shows a well-known limitation of compressed sensing (and machine learning) methodologies, namely that a good transferability cannot be expected \textit{a priori} for cases not previously encountered in the training or validation data. However, to put this in perspective, we note that this "poor" RMSE for the other facets is essentially of the same level as the RMSEs obtained for the widely used scaling relations in the first place.

In order to identify a descriptor that has a good compromise between accuracy for alloys and facets, we include 50\% of the facets data in the validation data (see Supplementary Fig. S5). The new best descriptor is found for the hyper parameters 8D, $\mathbf{\Phi}_1$. It is interesting to note that since this is a $\mathbf{\Phi}_1$ descriptor, the functional form of the features is much less complex than for the $\mathbf{\Phi}_3$ descriptor optimized for the alloys alone. This suggests that a less complex mathematical form is required for a descriptor that is transferable across both alloys and active site motifs.

The RMSEs (maxAEs) of the (8D, $\mathbf{\Phi}_1$) descriptor for the remaining 50\% of the facets data are found to be: (111): 0.17\,eV (0.44\,eV), (110): 0.21\,eV (0.59\,eV), (100): 0.24\,eV (0.48\,eV). These very moderate errors show that it is possible to identify a descriptor with a good predictive performance for a wide range of structural motifs as exemplified by the low-index fcc facets.
The improved performance on the facets data sets comes at the very moderate expense of increasing the RMSEs for the alloys to 0.18\,eV compared to 0.15\,eV before. We therefore suggest the (8D, $\mathbf{\Phi}_1$) SISSO descriptor for cases where simultaneous screening of alloys and a wide range of active site motifs is desired. It should be emphasized though that in general a good performance can only be expected for active sites types that resemble to some extent those types that the descriptor was optimized for.
Likely, more varied training and validation data will be required to identify a descriptor that would work for very different active site motifs such as kinks, vacancies and adatoms.

\subsection{Composition of descriptors}
Having confirmed the predictive performance of the identified descriptors, we now move on to discuss their composition. As an example of a descriptor identified by SISSO, we give in eq.\ \ref{eq:2D_desc} the best 2D descriptor of $\mathbf{\Phi}_3$ from Fig.\ \ref{fig:metals_RMSE}(a) (alloy validation data set):
\begin{equation}\label{eq:2D_desc}
\begin{split}
P^j = &\ c_1^j \cdot \left( \frac{V_{\rm ad}^2}{\rm PE} -  {\rm sin}({\rm site}_{\rm no}) + \frac{{\rm IP} \cdot {\rm sin}(\varepsilon_d)}{\varepsilon_d} \right) \\
+ &\ c_2^j \cdot \left( {\rm PE} \cdot f_d - \frac{f_{sp} \cdot {\rm log}({\rm DOS}_{sp}) \cdot (\varepsilon_d - W)}{V_{\rm ad}^2} \right) \\
+ &\ c_0^j \quad ,
\end{split}
\end{equation}
where $P^j$ is an adsorption energy of adsorbate $j$ and the primary features entering the descriptor are evaluated for the material/site combination relevant for $P^j$. The $d$-band center and $sp$-band properties such as the Pauling electronegativity are identified as highly important primary features. This was also found in previous studies employing artificial neural networks \cite{Ma2015,Li2017}, but is here expressed in an explicit non-linear functional form owing to the compressed sensing methodology. Among the remaining primary features entering the descriptor we especially highlight the DOS at the Fermi level (here of the $sp$-band), which is a feature not considered in these previous studies, even though its importance for the reactivity of TM surfaces was discussed already more than 30 years ago by Yang and Parr \cite{Yang1985}. 

An overview of the identified descriptors of each dimension and rung for both the alloy validation data set and the combined alloy and facets validation data set together with the fitting coefficients $c_i^j$ is given in Supplementary Table S8 and S9.  

\subsection{Enlarging the training data set}
The predictive performance of the identified descriptors for alloy screening is already impressive, given that no explicit information on alloys was given in the training data. However, a further advantage of data-driven approaches is that the learning can be systematically improved by enlarging the training data set. In contrast, the rigid format of linear scaling relations does not allow for significant improvements, even if fitting also to the alloys data, as evidenced by the scattering of the alloy data points around the fitted line in Fig.\ \ref{fig:scaling}(a) and (b).

To provide a simple estimate of the learning improvement possible when including also alloys in the training data, we identify a new SISSO descriptor (see Supplementary Table S10) based on the pooled metal and alloys data sets, but excluding the 23 DFT-calculated alloy data points (colored stars) shown in Fig.\ \ref{fig:scaling}(a) and (b). For this we use the hyperparameters that were found to be best for alloys (8D, $\mathbf{\Phi}_3$). The colored dots in Fig.\ \ref{fig:scaling}(a) and (b) show the SISSO predictions for the data points left out in the training. Already a visual inspection reveals that the agreement is very good. The maxAE (of 0.52\,eV) is found for the OH adsorption energy of the dark green point in Fig.\ \ref{fig:scaling}(a), which corresponds to OH adsorption on top of the Au atom in the Au@Ni alloy. Note that exactly this data point is also the maxAE (with the slightly larger value of 0.61\,eV) for the descriptor trained only on the pure metals. The RMSE over the 23 predicted alloy data points in Fig.\ \ref{fig:scaling}(a) and (b) decreases from 0.23\,eV (training on pure metals only) to 0.18\,eV (training on pooled data set).

Overall, the good agreement between the DFT-calculated values and the SISSO predictions shows that our models have the required accuracy to systematically search for outliers to scaling relations. In addition, our approach is computationally cheap enough to allow for the screening of immense alloy spaces. For this, we will next present a simple example.

\subsection{A first screening example}
In the following we make use of a SISSO descriptor (see Supplementary Table S11) identified using the hyperparameters (8D, $\mathbf{\Phi}_3$) and the entire pooled metals and alloys data sets for training. We predict the adsorption energies for the adsorbates and sites considered in Fig.\ \ref{fig:scaling}(a) and (b) on the additional 32 possible AB alloys (those that were not explicitly calculated by DFT) as shown with the gray dots. Similar to the explicitly DFT-calculated alloy data points, there is a considerable scatter around the scaling relation lines. This shows that there indeed exist many materials with potentially interesting catalytic properties, which would be missed by a scaling-relation-based screening approach.

A particularly interesting perspective for catalyst screening is to be able to search directly for candidate materials that break scaling relations. For many reactions, the incentive would be to break scaling relations in a desired way, since it has been suggested that scaling relations impose an upper limit to the possible catalyst activity. For example, it is known that for the oxygen evolution reaction it would be desirable to find a material where O is destabilized relative to OOH \cite{Rossmeisl2005}, and for electrochemical CO$_2$ reduction it is desirable to destabilize CO relative to CHO \cite{Li2016}. For other reactions, where optimum catalytic activity has hitherto been exclusively formulated in terms of singular descriptors, the interest would be to evaluate the effect of scatter in the binding of other important intermediates that hitherto has been assumed as fixed through scaling relations. For the oxygen reduction reaction (ORR), optimum catalytic activity has for instance been associated with an optimum oxygen adsorption energy \cite{Greeley2009}, while for selective ethanol synthesis both the carbon and the oxygen adsorption energy must be simultaneously optimized \cite{Medford2014b}. The activity at the top of this theoretical volcano curve is then independent of e.g.\ the OH (ORR) or CH (ethanol) adsorption energy, as the latter are connected to the optimum O or C adsorption energy through a scaling relation, respectively. 

In Fig.\ \ref{fig:scaling}(c) and (d) we specifically check on the scatter by showing histograms of the predicted adsorption energies for (c) O and OH and (d) C and CH at all potential adsorption sites of all 36 AB alloys. There are more predicted points for OH (around 1000) than for the other adsorbates (around 400), since OH adsorption can take place at more site types, i.e.\ also at top and bridge sites. The black shaded areas highlight in Fig.\ \ref{fig:scaling}(c) the ORR "volcano optimal" O adsorption energy on (111) facets \cite{Norskov2004} and in Fig.\ \ref{fig:scaling}(d) the optimal C adsorption energy on (211) facets for selective ethanol synthesis \cite{Medford2014b}. For this simple screening example, we will assume that only the most stable adsorption site of a given adsorbate plays a catalytic role, keeping in mind that in reality also less stable (meta-stable) sites could get populated at higher coverages. The SISSO approach directly gives us the energetics for the most stable and all meta-stable sites, so in general we are not limited to considering only most stable sites. Since the ORR volcano was developed for (111) facets, we search for materials for which the most stable O adsorption energy among the (111)-like (terrace) sites of the (211) facet falls within the desired range. This results in three candidate materials: PdPt, AgPd, and AgPt. The latter material is highlighted in Fig.\ \ref{fig:scaling}(c), since it has an OH adsorption energy on its most stable bridge1-$t$ site that is 0.23 eV lower than the value that would be predicted for this site from scaling relations (indicated by the black arrow) based on the O adsorption energy on its most stable hcp-$t$ site. The opposite behavior (a lower O adsorption energy relative to OH) is seen for the material shown with the green dot (PdPt). The cause of this breaking of scaling relations is thereby the slightly different oxygenate adsorption energy for Pt and Ag, and the fact that the preferred adsorption sites for OH and O have a different composition of Pt and Ag. A similar breaking of scaling relations is observed for the ethanol synthesis example. Here, SISSO recovers the scatter in the most stable CH adsorption energies for materials (RuAg, RhAu, RuAu, IrAu, and CuIr) that all have about the same most stable "optimum" C adsorption energy. This scatter shows the extent to which it is possible to tune the CH adsorption energy independently of the C adsorption energy and thereby further tailor the catalytic activity.

\subsection{A high-throughput screening perspective}
It should be noted that the moderate breaking of scaling relations observed in the simple examples from the previous section is related to the consideration of only a handful of "near-optimal" materials (out of a total of only 36 considered materials) and in particular the consideration of only most stable adsorption sites. A full assessment of the extent to which scaling relations can be broken on alloys with mixed-metal surfaces would only be revealed by a full high-throughput screening of hundreds of thousands of materials that is beyond the scope of the present study. Such a high-throughput screening could also involve the evaluation of a microkinetic model for each catalyst material that takes into account all possible adsorption sites of every adsorbate, as well as kinetic barriers for reaction and diffusion steps through BEP relations. If such a microkinetic model was initially carried out within the simplifying mean-field approximation \cite{Medford2015}, the cost of its evaluation would still only be a negligible fraction of the (already small) cost of carrying out a DFT calculation of the primary features of the clean catalyst surface for the descriptor evaluation. Once a selection of promising catalyst materials had been identified, a next step could then be the evaluation of a more thorough microkinetic model from e.g.\ a kinetic Monte Carlo simulation \cite{Andersen2017}, possibly taking into account also lateral interactions between the adsorbates through cluster expansion methods \cite{Stamatakis2016}. A full assessment of identified promising catalyst materials would ultimately also need to take into account other aspects such as bulk and surface segregation stability under realistic surface coverages for the chemical reaction and reaction conditions of interest, as well as stability against metal stripping in electrocatalysis applications.

\section{Conclusions}
In summary, we have used compressed sensing to identify new and better descriptors that allow to predict adsorption energies for a whole range of atoms and molecules at all potential surface sites of TMs and bimetallics formed of TMs. The descriptors can be obtained from a single DFT calculation of the clean surface and their predictive power extends over both multi-metallics and various surface facets. Importantly, this enables low-cost catalyst screening not only in materials, but also in active site space \cite{Reuter2017} with unprecedented accuracy. With respect to materials, the thereby enabled systematic identification and analysis of outliers to traditional scaling relation energetics seems particularly promising. With respect to active sites, the availability of energetic data for a wide range of site types paves the way to actively embrace the uncertainty in surface structure and composition of working catalysts.

\begin{suppinfo}
Supporting Information. Additional DFT and SISSO computational details, multi-task learning versus learning of separate descriptors, overview of largest prediction errors, all scaling relation plots, tested SIS values, hyperparameter testing with facets data in validation data set, and all identified descriptors. This material is available free of charge via the Internet at http://pubs.acs.org.
\end{suppinfo}

\begin{acknowledgement}
This project has received funding from the European Unions Horizon 2020 research and innovation program under grant agreement No. 676580, The NOMAD Laboratory, a European Center of Excellence. The authors also gratefully acknowledge the Gauss Centre for Supercomputing e.V. (www.gauss-centre.eu) for funding this project by providing computing time on the GCS Supercomputer SuperMUC at Leibniz Supercomputing Centre (www.lrz.de). We also thank Luca Ghiringhelli for helpful discussions and a careful proofreading of the manuscript. Runhai Ouyang has written the multi-task SISSO code \cite{Ouyang2018} used in the present work and provided technical assistance.
\end{acknowledgement}

\bibliography{Andersen_beyond_scaling_relations_references}

\end{document}